\pdfoutput=1
 \documentclass[fleqn,usenatbib,usedcolumn]{mnras}
 \usepackage{newtxtext,newtxmath}
\usepackage[T1]{fontenc}
 \usepackage{graphicx}
 \let\la=\lesssim 
\usepackage{natbib,hyperref}
 \usepackage{epstopdf}
 \hypersetup{pdfauthor={S. J. Morrison},
               pdftitle={Gap Formation in Planetesimal Disks Via Divergently Migrating Planets},
               pdfkeywords={methods: miscellaneous -- minor planets, asteroids: general -- planet-disc interactions -- circumstellar matter -- planetary systems},
               bookmarksnumbered=true}
 \newcommand{\Rh}{R_{\rm Hill}}

\title[Gaps via Migrating Planets]{Gap Formation in Planetesimal Disks Via Divergently Migrating Planets}

\author[S. J. Morrison et al.]{
Sarah J. Morrison,$^{1}$\thanks{E-mail: smorrison@psu.edu}
Kaitlin M. Kratter$^{2}$
\\
$^{1}$Center for Exoplanets and Habitable Worlds, 525 Davey Laboratory, The Pennsylvania State University, University Park, PA, 16802, USA\\
$^{2}$Steward Observatory,University of Arizona, 933 North Cherry Avenue, Tucson, AZ 85721, USA
}

\date{Accepted 2018 September 25. Received 2018 September 25; in original form 2018 January 28}

\pubyear{2018}
\begin{document}

 \label{firstpage}
\pagerange{\pageref{firstpage}--\pageref{lastpage}}
\maketitle

\begin{abstract} 
While many observed debris disks are thought to have gaps suggestive of the presence of planets, direct imaging surveys do not find many high mass planets in these systems. We investigate if divergent migration is a viable mechanism for forming gaps in young debris disks with planets of low enough mass to currently elude detection. We perform numerical integrations of planet pairs embedded in planetesimal disks to assess the conditions for which divergent, planetesimal-driven migration occurs and gaps form within the disk. Gap widths and the migration rate of planets within a pair depend on both disk mass and the degree to which the planets share disk material. We find that planet pairs with planets more massive than Neptune can produce gaps with widths similar to their orbit distance within 10 Myr at orbit separations probed by direct imaging campaigns. Pairs of migrating super-Earths likely cannot form observable gaps on the same time and distance scales, however. Inferring the responsible planet masses from these gaps while neglecting migration could overestimate the mass of planets by more than an order of magnitude. 
\end{abstract}

\section{Introduction}

With improving direct imaging capabilities, we are now gaining the ability to detect massive planets interacting with debris disks on distance scales similar to the outer Solar System. Meanwhile, the sample of debris disks known to possess wide gaps also continues to grow. Similar to our Solar System, these gaps can be wide with an outer to inner debris belt distance ratio of $\sim$10 \citep[e.g.][]{Su:2013, Kennedy:2014, Su:2015}. For unresolved debris disks, these gapped systems require multiple dust thermal emission temperatures to fit the system's spectral energy distribution (SED) \citep{Backman:2009, Chen:2009, Morales:2009, Ballering:2013}. For most of these systems with A-type host stars and/or with far-IR/mm detections, the presence of gaps inferred from SEDs is robust against the alternative interpretation of a single debris belt with a range of dust temperatures arising from grain size differences \citep{Kennedy:2014}. Dust grains in debris disks are subject to radiation pressure and Poynting-Robertson drag, which cause dust particles to be either blown out or to drift inward on short timescales ($<10^4$ years), so the existence of a gapped debris disk implies dynamical stirring of leftover planetesimals to produce dust, and clearing of inwardly drifting dust by planets to maintain a gap \citep{Wyatt:2008}. Consequently, direct imaging surveys have targeted young (10s Myr) debris disk systems in particular and are sensitive to multi-Jupiter mass planets at distances of 10s of AU \citep{Nielsen:2013, Meshkat:2015, Bowler:2016}. Yet, for giant planets to be responsible for inferred debris disk gaps, these surveys should be detecting more planets than they actually do \citep{Bowler:2016, Morrison:2016}. However, reanalysis of some direct imaging data might substantially reduce the discrepancy (Stone et al. in prep). 

Planet occurrence rates from radial velocity and transit surveys suggest that lower mass planets are more common \citep{Fressin:2013}, but planets $\lesssim1M_{\rm{Jupiter}}$ are not currently observable by direct imaging surveys \citep{Bowler:2016}. Moreover, a system containing planets similar to our outer Solar System's would not be currently observable. The current debris populations of our Solar System suggest that the outer planets likely started in a more compact configuration and have since migrated apart as they scattered planetesimals from the early asteroid and Edgeworth-Kuiper debris belts \citep[e.g.][]{Fernandez:1984, Malhotra:1993, Murray-Clay:2005, Minton:2009, DeMeo:2014}. In particular, Neptune must have migrated outward in order to reproduce the resonant objects in the Edgeworth-Kuiper belt, and this outward migration was accomplished as Neptune exchanged angular momentum with the residual disk in its vicinity during the scattering process \citep[e.g.][]{Malhotra:1995, Hahn:1999}. Here we investigate the degree to which divergent planet migration could plausibly form gaps in debris disks, and its implications for inferring planetary system architectures from disk observations. 
\subsection{Previous work on planetesimal-driven planet migration}
For a single planet migrating in a gas-dominated disk (like a protoplanetary disk) or a gas-less planetesimal disk (like a massive debris disk), previous works have analytically and numerically estimated the angular momentum exchange between the disk and planet  to obtain the planet's migration rate \citep{Fernandez:1984, Ida:2000, Kirsh:2009}. These migration rates depend on the properties of the disk near the planet (such as the local surface density) and the nature of the encounter between the disk material and the planet (i.e. scattering or accretion dominated). \citet{Kirsh:2009} found that single planets undergoing planetesimal-driven migration typically migrate inward due to the shorter conjunction timescale with disk material interior rather than exterior to the planet's orbit. Also, the migration rate of a single planet depends more strongly on disk mass when planets are $\lesssim$10x the mass of the local disk material. 

The migration of more than one planet becomes more complicated, however. For most disk surface density profiles, the mass ratio between the planet and the locally available disk material will differ with orbital separation even for equal mass planets, so migration rates can differ for each planet within a multi-planet system. Additionally, the disk material between more closely separated planet pairs will be perturbed by both planets, resulting in different encounter timescales and effective planet migration rates than predicted for single planets alone. When planets undergo divergent migration, their orbit spacings increase and they can also `hop' over mean motion resonances with respect to each other, which will cause a sudden change in the planets' eccentricities and produce a response in the disk. Because of all of these variables, previous studies of divergent migration, including models of the migration history of the outer Solar System \citep[e.g.][]{Gomes:2005}, have been difficult to generalize and provide insight into the broader context of the evolutionary history of observed debris disk systems.
\section{Methods}\label{S:Methods}
In an effort to investigate the mechanics of multi-planet migration in a generalized fashion, we explore the impact of disk mass and planet architecture on planet migration rates and gap opening timescales relevant for observed debris disk systems. We consider two planets at a given separation embedded within a disk of massive planetesimals. We characterize the disk interacting with the planets via the mass ratio between disk material local to the planet and the planet itself, $M_d\equiv M_{\rm {local}}/M_{\rm{p}}$. The angular momentum exchange from encounters between nearby material and the planet rather than the total disk extent determines the rate of the planet's migration on short timescales. \citet{Kirsh:2009} determined that this local source of planetesimals drives single planet migration. The size of the relevant region is a few times the planet's Hill radius,
\begin{equation}\label{eq:Rhm}
\Rh=X a_p=\left(\frac{M_p}{3M_*}\right)^{1/3}a_p.
\end{equation}
In this paper, we adopt $M_{\rm{local}}$ to be the mass of the disk enclosed within a planet's outer encounter zone, here defined as 2.5$\Rh$ outside the planet's orbit.

\subsection{Numerical Simulations of Migrating Planet Pairs}
We numerically simulated systems of two planets with individual planet-star mass ratios, $M_p/M_*$, of $10^{-3}$, $10^{-4}$, or $10^{-5}$ embedded in disks with dimensionless disk mass $M_d$ of 1/3, 1/10, 1/30 or 1/100 measured relative to the outer planet. For a solar mass host star, these planet masses correspond to $\sim1 M_{\rm{Jupiter}}$, $2 M_{\rm{Neptune}}$, and $3M_\oplus$, respectively. Unless otherwise noted, planets were initialized on circular orbits such that $a_{\rm{outer}}=1.5a_{\rm{inner}}$, which corresponds to a period ratio of $\sim$1.84. We show in Section \ref{S:ictranslate} how to translate between these dimensionless, initial mass and distance quantities to absolute masses and distances when applying these results to particular systems. As in \citet{Kirsh:2009}, we set initial disk particle eccentricities and inclinations to be Rayleigh distributed about a value of 0.01 with inclinations twice this value in radians. This corresponds to a dynamically relaxed disk \citep{Tremaine:1998} under conditions that typically follow the runaway growth process in planet formation \citep{Ida:1992a}. Disks had a surface density profile $\Sigma\propto a^{-1}$ and contained at least 10,000 particles with particle masses chosen to achieve a given $M_{\rm{local}}$. A summary of the simulations we performed are shown in Table~\ref{t:sims}.

\begin{table}
\centering
\begin{tabular}{ccccc}
$M_{\rm outer}/M_*$ & \multicolumn{1}{l}{$M_{\rm inner}/M_*$} & $M_{\rm {local}}/M_{\rm {outer}}$         & \multicolumn{1}{l}{$N_{\rm {disk}}$} & \multicolumn{1}{l}{Disk Extent} \\ \hline 
$10^{-3}$           & $10^{-3}$                               & $\frac{1}{30},\frac{1}{100}$              & $10^4$                               & 0.1-10                          \\
$10^{-3}$           & $10^{-3}$                               & $\frac{1}{3},\frac{1}{10}$                & $10^4$                               & 0.1-5                           \\
$10^{-3}$           & $10^{-4}$                               & $\frac{1}{100}$                           & $10^4$                               & 0.1-10                          \\
$10^{-3}$           & $10^{-4}$                               & $\frac{1}{10}, \frac{1}{30}$              & $10^4$                               & 0.1-5                           \\
$10^{-3}$           & $10^{-4}$                               & $\frac{1}{3}$                             & $2\times10^4$                        & 0.1-5                           \\
$10^{-4}$           & $10^{-3}$                               & $\frac{1}{100}$                           & $10^4$                               & 0.1-10                          \\
$10^{-4}$           & $10^{-3}$                               & $\frac{1}{10}, \frac{1}{30}$              & $10^4$                               & 0.1-5                           \\
$10^{-4}$           & $10^{-3}$                               & $\frac{1}{3}$                             & $2\times10^4$                        & 0.1-5                           \\
$10^{-4}$           & $10^{-4}$                               & $\frac{1}{100}$                           & $10^4$                               & 0.1-10                          \\
$10^{-4}$           & $10^{-4}$                               & $\frac{1}{10}, \frac{1}{30}$               & $10^4$                               & 0.1-10                          \\
$10^{-4}$           & $10^{-4}$                               & $\frac{1}{3}$                             & $2\times10^4$                        & 0.1-5                           \\
$10^{-4}$           & $10^{-5}$                               & $\frac{1}{30}, \frac{1}{100}$             & $10^4$                               & 0.1-5                           \\
$10^{-4}$           & $10^{-5}$                               & $\frac{1}{3}, \frac{1}{10}$               & $2\times10^4$                        & 0.1-5                           \\
$10^{-5}$           & $10^{-3}$                               & $\frac{1}{100}$                           & $10^4$                               & 0.1-5                           \\
$10^{-5}$           & $10^{-3}$                               & $\frac{1}{3}, \frac{1}{10}$, $\frac{1}{30}$ & $2\times10^4$                        & 0.1-5                           \\
$10^{-5}$           & $10^{-4}$                               & $\frac{1}{100}$                           & $10^4$                               & 0.1-5                           \\
$10^{-5}$           & $10^{-4}$                               & $\frac{1}{3}$, $\frac{1}{10}$, $\frac{1}{30}$ & $2\times10^4$                        & 0.1-5                           \\
$10^{-5}$           & $10^{-5}$                               & $\frac{1}{100}$                           & $10^4$                               & 0.1-5                           \\
$10^{-5}$           & $10^{-5}$                               & $\frac{1}{3}$, $\frac{1}{10}$, $\frac{1}{30}$ & $2\times10^4$                        & 0.1-5                          
\end{tabular}
\caption{Disk and planet parameters for simulations performed. Disk extent measured in units of initial outer planet orbit distance. $N_{\rm{disk}}$ is the number of disk particles and each case of local disk-outer planet mass ratio ($M_{\rm {local}}/M_{\rm {outer}}$) per row is a separate simulation with the given planet pair mass combination.}
\label{t:sims}
\end{table}
 Within the simulations, we model two types of objects orbiting the host star: planets and disk particles. The planets experience the gravity of disk particles and gain mass as disk particles collide with the planet, but there is no disk self-gravity or collisions between disk particles. Orbital eccentricities of planetesimals in a planet's encounter zone are excited by the planet faster than collisions can damp them \citep{Bryden:2000, Ida:2000}, so we neglect particle-particle collisions to reduce simulation runtimes. We use a modified version of the REBOUND HERMES integrator for all numerical integrations, which is a hybrid integrator that switches between using a symplectic Wisdom-Holman mapping method, WHFAST \citep{Rein:2015b}, for particles distant to the planets and a Gauss-Radau method (IAS15; \citet{Rein:2015a}) for particles closer to the planets. This built-in switch is triggered if particles get within 4 Hill radii of a planet. For improved energy conservation, we added a second trigger to switch to IAS15 when the particle-planet encounter timescale, determined by the particle's velocity relative to the planet, came within one fifth of the WHFAST integration timestep. Typical fractional energy changes were less than $4\times10^{-11}$ over $10^4$ orbits. We validated this modified integrator by successfully replicating the orbit migration and mass growth of individual $18M_{\oplus}$ and $27M_{\oplus}$ planets with the same disk parameters reported in \citet{Kirsh:2009}.

Our default disks have 10,000 particles spanning 0.1$a_{\rm{outer}}$-10$a_{\rm{outer}}$. To avoid introducing artificial stochasticity in the planet's migration from individual planet-particle encounters, we sought to kept the mass ratio between a planet and disk particle to be $\lesssim10^{-4}$. Consequently, we increased the particle number and shortened the outer disk extent to 5$a_{\rm{outer}}$ in our simulations of lower mass planets in higher mass disks. From comparing trial simulations with the same $M_{\rm{local}}$ and planet configurations but different disk particle resolutions, we find migration outcomes to be the same below the $\lesssim10^{-4}$ planet-to-disk particle mass ratio threshold. \citet{Kirsh:2009} also reported artifacts in a planet's migration when disk particles exceeded $\sim$1/600th of the migrating planet's mass for the same disk surface density power law scaling. 

We integrated all systems for at least $10^{4.5}$ orbits of the outermost planet. We do not want disk edge effects to contaminate our investigations on migration and gap opening, and therefore do not include simulations in the following analysis beyond times for which one of the planets reaches the edge of the disk. This occurred only for one simulation, which contained: an inner planet with $M_{\rm {inner}}/M_{*}$=10$^{-3}$, an outer planet with $M_{\rm {outer}}/M_{*}$=10$^{-4}$, and a disk such that $M_{\rm {local}}/M_{\rm {outer}}$=1/3.

 \section{Results}\label{S:Results}
\begin{figure}
\centering
\includegraphics[width=\columnwidth]{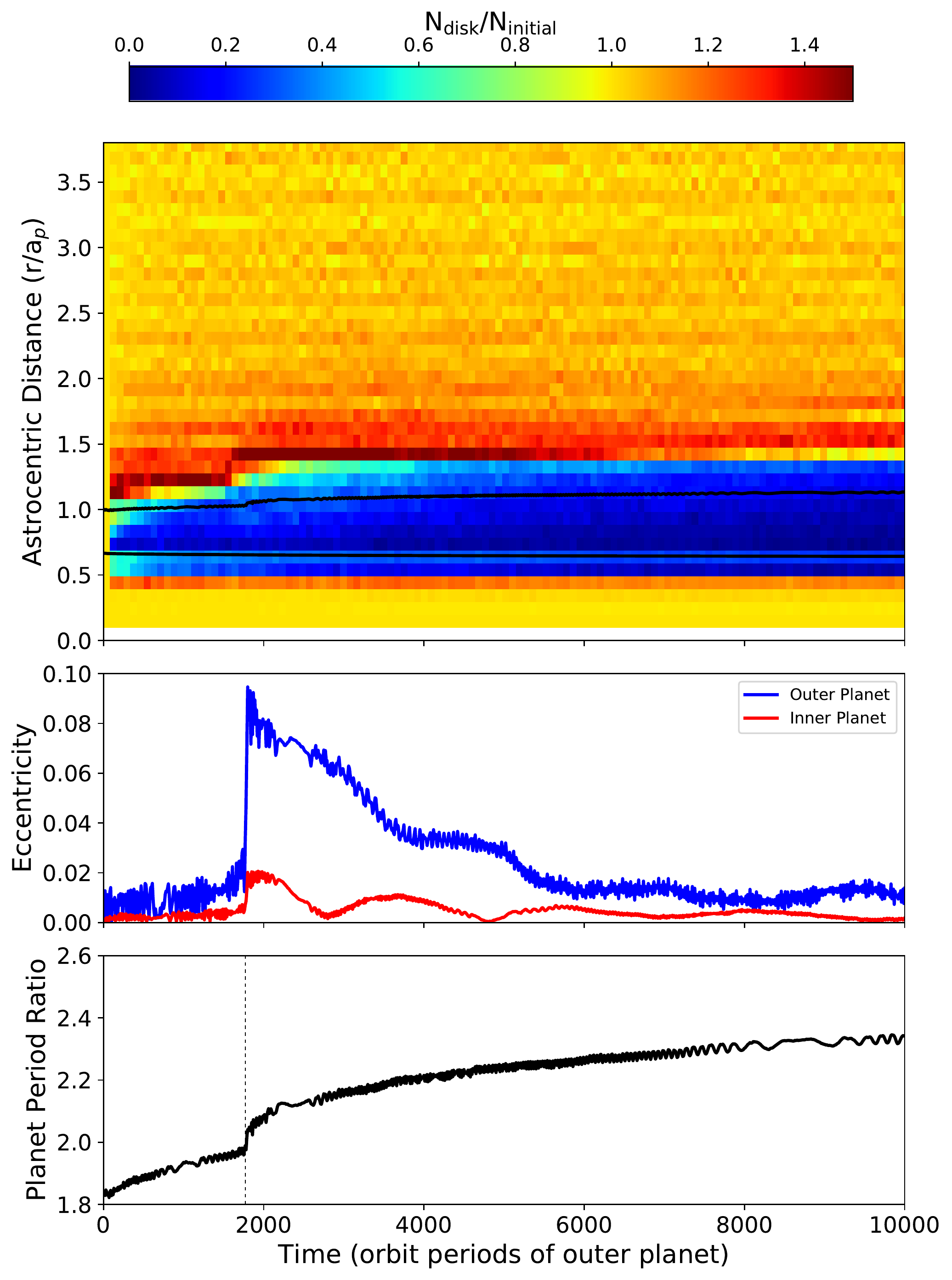}
\caption{A representative example of the disk and planet orbit evolution. The inner planet has a mass ratio of $M_p/M_*=10^{-3}$ and the outer planet has a mass ratio of $M_p/M_*=10^{-4}$ in a $M_{\rm{local}}/M_{\rm{outer}}=0.1$ disk. Top plot shows disk number density at a given time-averaged orbit distance compared to the initial number density. Solid black lines show the semimajor axis of each planet. Bottom two plots show the eccentricity and period ratio evolution of the planet pair over time.}
\label{fig:tevoex}
\end{figure}

 In nearly all simulations, planets undergo divergent migration and open gaps. As an example of this, we show the time evolution of the disk and planets in Figure~\ref{fig:tevoex} for the case with $M_{\rm {inner}}/M_{*}$=10$^{-3}$, $M_{\rm {outer}}/M_{*}$=10$^{-4}$, and $M_{\rm {local}}/M_{\rm {outer}}$=1/30. To quantify the time evolution of the disk surface density, we use the time-averaged radial distance of disk particles in their osculating orbits, $r_{\rm{ave}}=a_{\rm{particle}}(1+0.5e^2_{\rm{particle}}$). We count the number of particles within a given $r_{\rm{ave}}$ bin compared to the original population in that binned annulus. The widths of these bins were 10\% of the outer planet's initial orbit distance. This ensured that the annuli had sufficient disk particle numbers initially to track broad trends in disk depletion or enhancements rather than more artificial, stochastic records of depletions/enhancements due to orbit evolution of individual disk particles. In this example case, the planets rapidly deplete the disk between them and cause the gap to grow in extent as the outer planet migrated outward and the inner planet migrated inward. We now examine the degree of planet migration, gap formation, and their dependences in further detail for the full suite of planet pair and disk combinations.

Planet migration rates in multi-planet debris disk systems differ from single-planet migration rates when planets exchange disk material. The degree to which neighboring planets can exchange disk material depends on whether material scattered by one planet can encounter the other. We quantify this for each planet by deriving the critical eccentricity for planetesimals in the disk from the encounter zone of one planet to cross the encounter zone of the other planet. The encounter zones are the annuli interior and exterior to the planet's orbit defined in \S~\ref{S:Methods} for which local disk material directly interacts with the planet. A particle initially at the boundary of the outer planet's interior encounter zone has a semimajor axis of $a_2(1-CX_2)$ where $C=2.5$ and the subscript 2 refers to the outer planet's properties and $X$ is the Hill factor from Equation \ref{eq:Rhm}. The pericenter of that particle is $a_2(1-CX_2)(1-e_{\rm{particle}})$ where $e_{\rm{particle}}$ is the particle's eccentricity. To first approximation, for this particle to also have encounters with the inner planet, its orbit should cross the inner planet's outer encounter zone. This condition for sharing is:
\begin{equation}\label{eq:s}
a_1(1+CX_1)\gtrsim a_2(1-CX_2)(1-e_{\rm{particle}}).
\end{equation}
We define the ratio of the right and left sides of Equation~\ref{eq:s} as the `sharing ratio' with respect to the outer planet
\begin{equation}\label{eq:Sratio2}
S_{\rm{outer}}=\frac{a_1(1+CX_1)}{a_2(1-CX_2)(1-e_{\rm{particle}})}.
\end{equation}
Analogously for a particle located at the boundary of inner planet's outer encounter zone, the sharing ratio with respect to the inner planet is then
\begin{equation}\label{eq:Sratio1}
S_{\rm{inner}}=\frac{a_1(1+CX_1)(1+e_{\rm{particle}})}{a_2(1-CX_2)}.
\end{equation}
The higher this ratio, the greater the degree of exchange of disk material between the planets. We define the critical eccentricity, $e_{\rm{crit}}$, to be the eccentricity of the particle necessary for the sharing ratio to equal to 1. The critical eccentricity for a particle at the inner encounter zone of the outer planet is
\begin{equation}\label{eq:ecrit2}
e_{\rm{crit, outer}}=1-\frac{a_1(1+CX_1)}{a_2(1-CX_2)}
\end{equation}
and by analogy, the critical eccentricity for a particle at the outer encounter zone of the inner planet is
\begin{equation}\label{eq:ecrit1}
e_{\rm{crit, inner}}=\frac{a_2(1-CX_2)}{a_1(1+CX_1)}-1.
\end{equation}
Typical eccentricities for particles interacting with a nearby planet will be roughly the Hill factor, $e_{\mathrm{Hill}}\equiv X$, ranging from 0.01 to 0.07 for $M_p/M_*=10^{-5}$ and $10^{-3}$, respectively. We assess the capability of one planet to share disk material with the other by considering the ratio between $e_{\mathrm{Hill}}$ and $e_{\mathrm{crit}}$. In subsequent sections, we show that this sharing partially accounts for the variation in migration rates and resulting gap widths arising from the architecture of planet pairs. 

 \subsection{Migration Rate Dependences}\label{S:R:migrates}
\begin{figure}
\centering
\includegraphics[width=0.95\columnwidth]{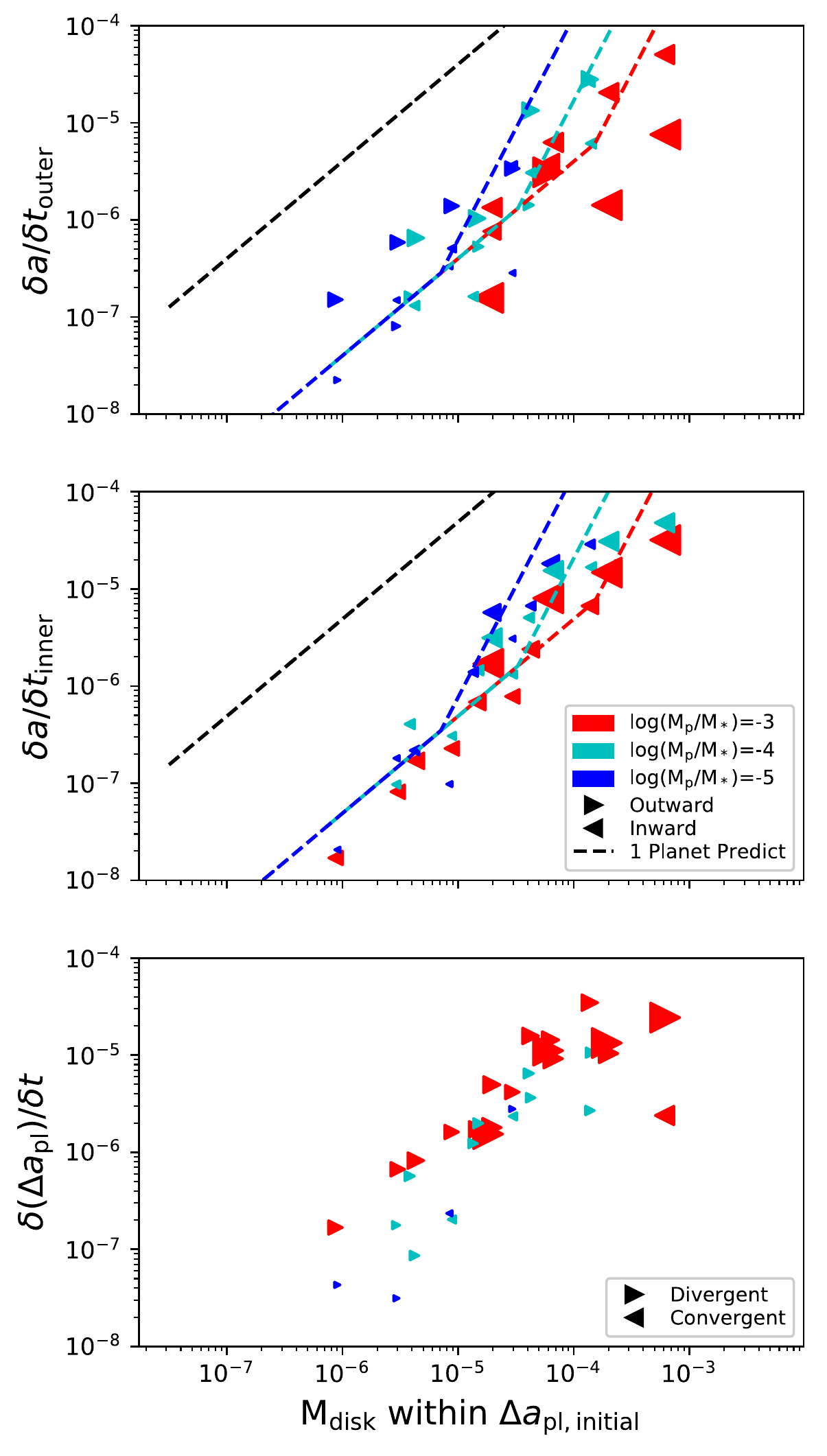}
\caption{Outer and inner planet migration rates (top and middle plots) and change in planet separation (bottom plot) versus initial disk mass ($M_{\rm{disk}}$) between the planets over $10^4$ outer planet orbits. Distances and times for calculated rates are reported with reference to the initial semi-major axis and orbital period of the outer planet. $M_{\rm{disk}}$ is normalized to $M_*$. Dashed lines indicate predictions of $|\rm{d}a/\rm{d}t|$ for single, isolated planets from semi-empirical estimates by \citet{Ida:2000} (black) and Eq 16 in \citet{Kirsh:2009} (colors). For all subplots, symbol size denotes degree of sharing of planetesimals between the two planets, defined by the ratio with respect to the highest mass planet of the Hill eccentricity to threshold eccentricity for crossing the other planet's encounter zone (Equations \ref{eq:ecrit2} and \ref{eq:ecrit1}). The smallest symbols correspond to a ratio of $\sim$0.05 and the largest to $\sim$1. Color indicates the outer planet mass for the top plot, and  inner planet mass for the middle plot, and highest planet mass within a simulated pair for for the bottom plot.}
\label{fig:migratesMdisk}
\end{figure}

Single planet migration is a strong function of disk mass. We also find this trend holds for multi-planet systems. For planet pairs, we calculate the migration rate of the individual planets and the `joint migration' rate by measuring the change in semimajor axis and the change in planet-planet separation, respectively, over the timescale of interest. All distances are scaled to the initial $a_{\rm{outer}}$. We show in Figure~\ref{fig:migratesMdisk} that the individual and joint migration rates over $10^4$ orbits are faster when the initial disk mass between the planets is higher. The migration rate of each planet follows a power law trend with disk mass as expected from single planet migration \citep{Kirsh:2009}, albeit with greater scatter for a given disk mass. The magnitude and overall direction of the migration also differs between the migration of two closely separated planets versus a single planet alone. The migration rate of each planet is about two orders of magnitude lower than the rate expected if the planet were exchanging angular momentum with the local disk material during a single scattering event \citep{Ida:2000}. However, the empirical fits by \citet{Kirsh:2009} from inwardly migrating single planets in low mass disks come closer to approximating the magnitude of the individual planet migration rates in our simulations over the timescales we feature in this work. Over longer timescales, individual planets within pairs migrate more slowly than if they were single planets. Over time, their migration slows as planets deplete the disk material between them, exchange less disk material as they migrate apart, and are left with less local disk mass on only one side of their orbit. As in the single planet case, at fixed disk mass, low mass planets migrate faster than higher mass planets. This trend holds for each planet in our multi-planet simulations, but with additional scatter dependent on the sharing ratio.

\begin{figure}
\centering
\includegraphics[width=\columnwidth]{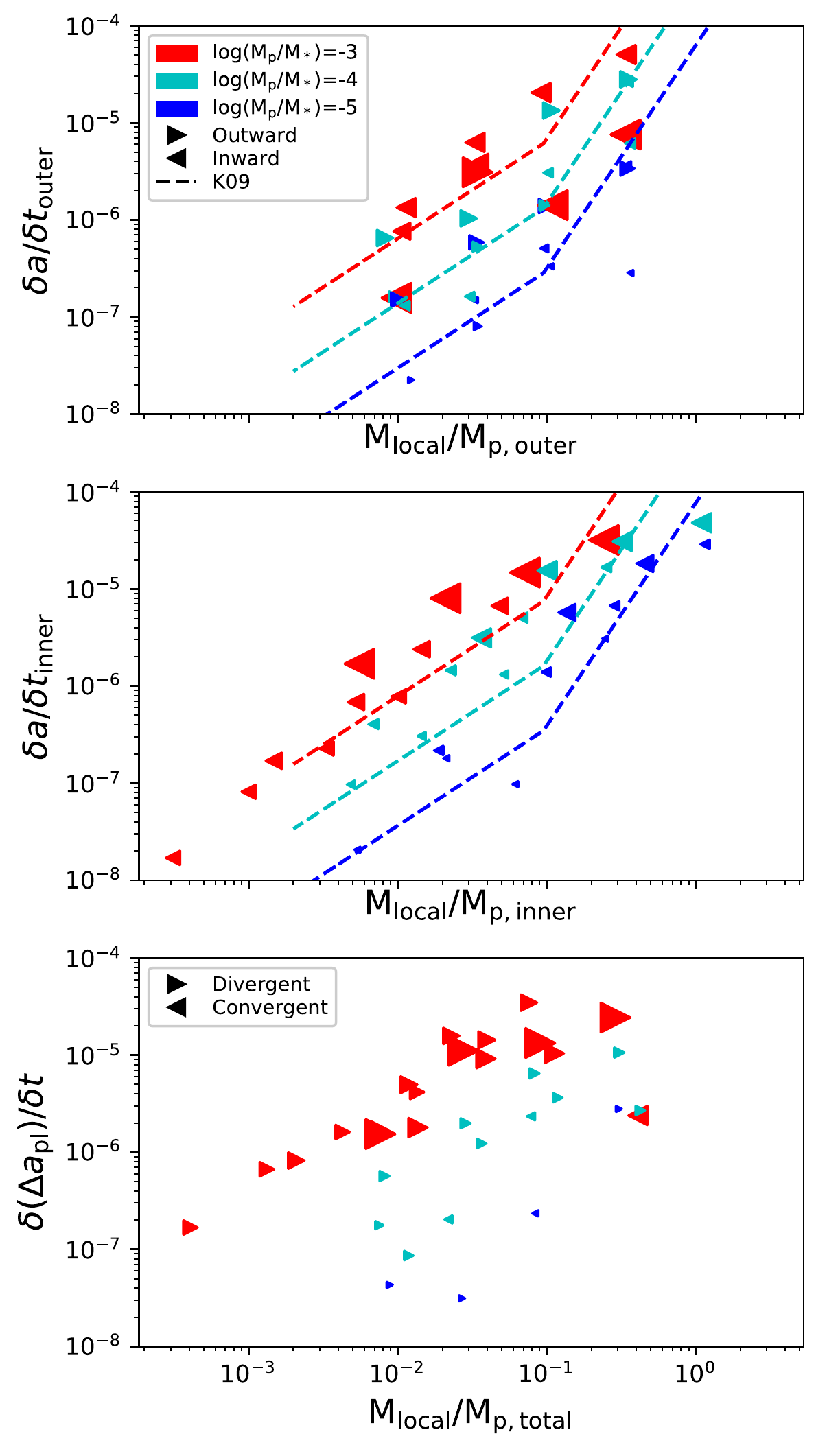}
\caption{Migration rates and change in planet separation versus local relative disk-planet mass ratio over $10^4$ outer planet orbits. Color delineates the outer planet mass for the top plot, inner planet mass for the middle plot, and highest planet mass within a simulated pair in the bottom plot. Dashed lines show the planet mass dependent empirical fit of migration rate magnitude from the inward migration rates of single, isolated planets from \citet{Kirsh:2009}.}
\label{fig:migratesMrel}
\end{figure}

To account for the differences in migration rates within similar disks due to planet architecture, we plot the migration rates and change in planet separation over $10^4$ orbits as a function of initial local disk to planet mass ratio in Figure~\ref{fig:migratesMrel}. Even for a given local disk-planet mass ratio, planet migration rates can be more than an order of magnitude different for different planetary system architectures. The inner planet migrates faster for the same disk mass if it shares more disk material with the outer planet (shown in the symbol sizes of the middle plot in Figure~\ref{fig:migratesMrel}). Material exterior to its orbit that is shared more readily with the outer planet does not encounter the inner planet again on timescales short enough to balance the angular momentum exchange with interior disk material. Therefore, the inner planet migrates inward faster. Analogously, outer planets that share more material also typically migrate faster for the same disk mass with the exception of pairs that contain $M_p/M_*=10^{-3}$ planets. This exception is likely due to the scattering efficiency and rapid scattering timescale of disk material by massive planets. In closely separated planet pairs, a $M_p/M_*=10^{-3}$ inner planet starves a $M_p/M_*=10^{-3}$ outer planet of some disk material interior to its orbit, slowing the outer planet's inward migration while insufficiently depleting the disk material between the planets to cause outward migration of the outer planet.

 \subsection{Conditions for Divergent Migration and the Growth of Gaps}\label{S:R:divmig}
With an understanding of what influences individual planet migration rates, now we examine the pair of planets together to discuss what influences the change in planet-planet spacing with time. Divergently migrating planets, whose separation grows as they migrate, would facilitate the growth of gaps in debris disks. The spacing between planets will increase in the following scenarios: 1) the inner planet migrates inward while the outer planet migrates outward, 2) the outer planet migrates inward more slowly than the inner planet, or 3) the inner planet migrates outward more slowly than the outer planet. In practice, the inner planet always migrated inward in our simulations so scenario 3 never occurred. For most of our simulations, the planets did, in fact, divergently migrate as evidenced in the bottom plots of Figures~\ref{fig:migratesMdisk} and \ref{fig:migratesMrel}. For lower disk masses and lower mass planet pairs with low sharing ratios (particularly for pairs containing a $M_p/M_*=10^{-5}$ planet), scenario 2 operated. Scenario 2 also operated for pairs containing $M_p/M_*=10^{-3}$ outer planets. For pairs with outer planets $M_p/M_*=10^{-4}$ and $10^{-5}$, the planets divergently migrated under scenario 1 when paired with a higher mass inner planet, since the inner planet could perturb more disk material out of the outer planet's inner encounter zone such that encounters with the outer disk dominated the outer planet-disk angular momentum exchange.

For the majority of cases where the spacing between planets increased, corresponding gaps within the disk also grew. Steady gap growth was punctuated by rapid increases as the planets `hopped' major mean motion resonances with respect to each other. A representative example of this behavior is seen in  Figure~\ref{fig:tevoex}. Since the planets in this investigation started at a period ratio of 1.84, the 2:1 resonance was the first major resonance encountered as the planets divergently migrated. As the two planets approached the resonance, their eccentricities were excited and then later damped by the disk. During this period of eccentricity excitation, the disk gap jumped in size up to tens of percent as the planets interacted with more disk material. Of our simulated pairs, the magnitude of this jump in gap width was most pronounced for planet pairs with a $M_p/M_*=10^{-3}$ and $M_p/M_*=10^{-4}$ mass combination. 

\subsection{Gap Depletion and Extent}\label{s:gapw}
\begin{figure}
\centering
\includegraphics[width=\columnwidth]{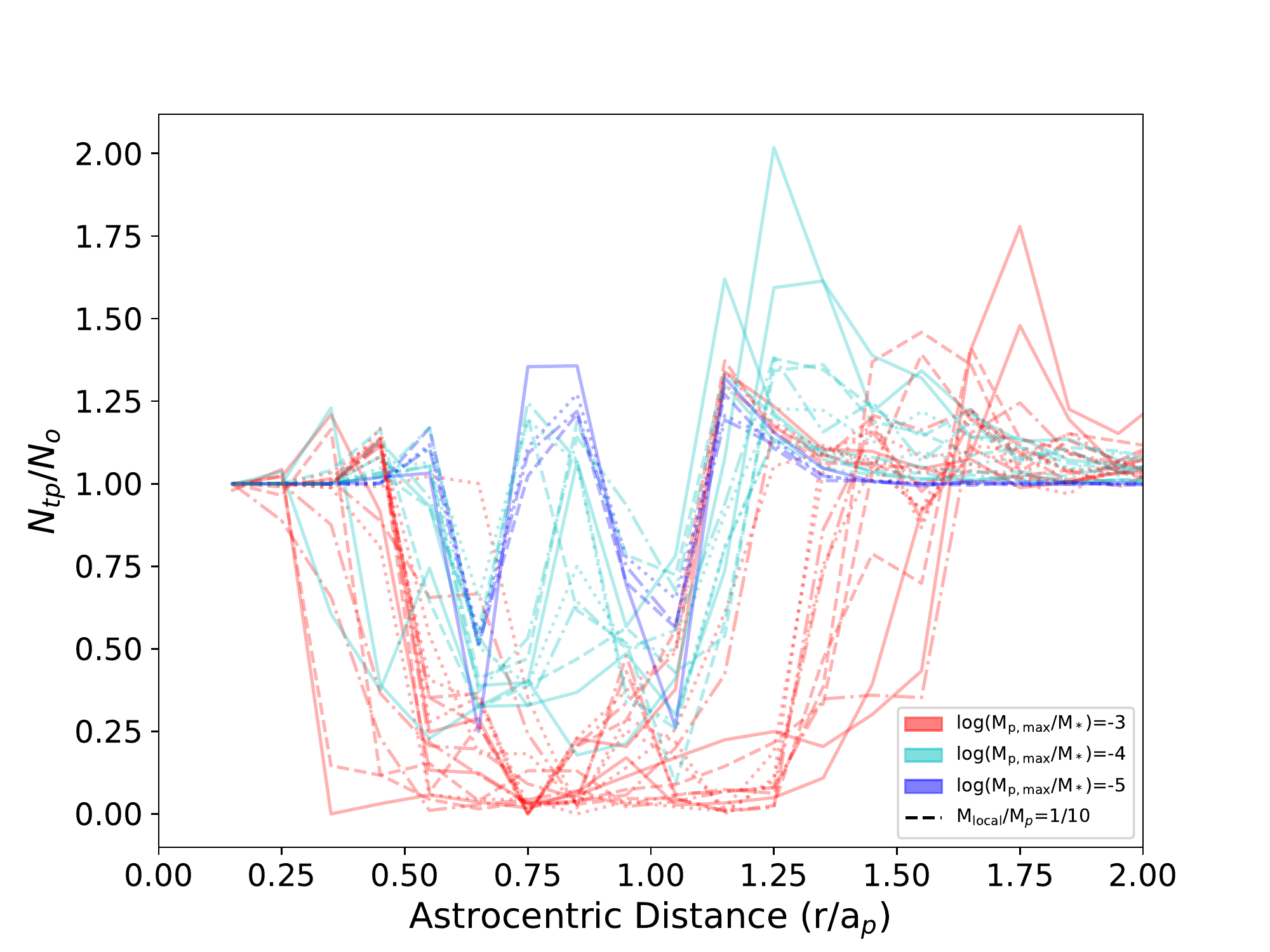}
\caption{Radial profile of the fraction of disk particles within a given $r_{\rm{ave}}$ bin after $10^4$ outer planet orbits compared to initially. Color delineates the highest initial planet mass within the pair. Line style denotes disk mass (in terms of $M_{\rm {local}}/M_{\rm {outer}}$).}
\label{fig:gapdep}
\end{figure}

In an effort to determine the relationships between gaps and the migrating planets responsible, first we consider how to define a gap. In radial disk profiles from spatially resolved disk observations, gaps are typically identified as depletions relative to a nearby area of the disk. Here we consider gaps as depletions relative to the amount of material that was originally in the radial annulus. We chose this approach since all of our disks started with the same radial profile. In Figure~\ref{fig:gapdep}, we show the radial profile of the fraction of particles after $10^4$ orbits.   Pairs containing higher planet masses create wider and more depleted gaps than lower mass planet pairs. Lower mass planet pairs also do not as fully deplete the material between them, but, in fact, produce concentrations of material between them. The outer edge of the gap in most cases also includes a more pronounced enhancement of material as some disk material is swept into mean motion resonances as the planets migrate.

In subsequent figures and analyses, we report the width of the overall gap, d$_{\rm{gap}}$, based on the minimum and maximum radial distance depleted by >50\%. Given the steepness of the gap `walls', the relationships we describe next do not change significantly with choosing a higher level of depletion to define a gap. We categorize these gaps as `depleted' or `non-depleted' based on whether or not the entire region between these boundaries is also depleted by >50\%. 

\begin{figure}
\centering
\includegraphics[width=\columnwidth]{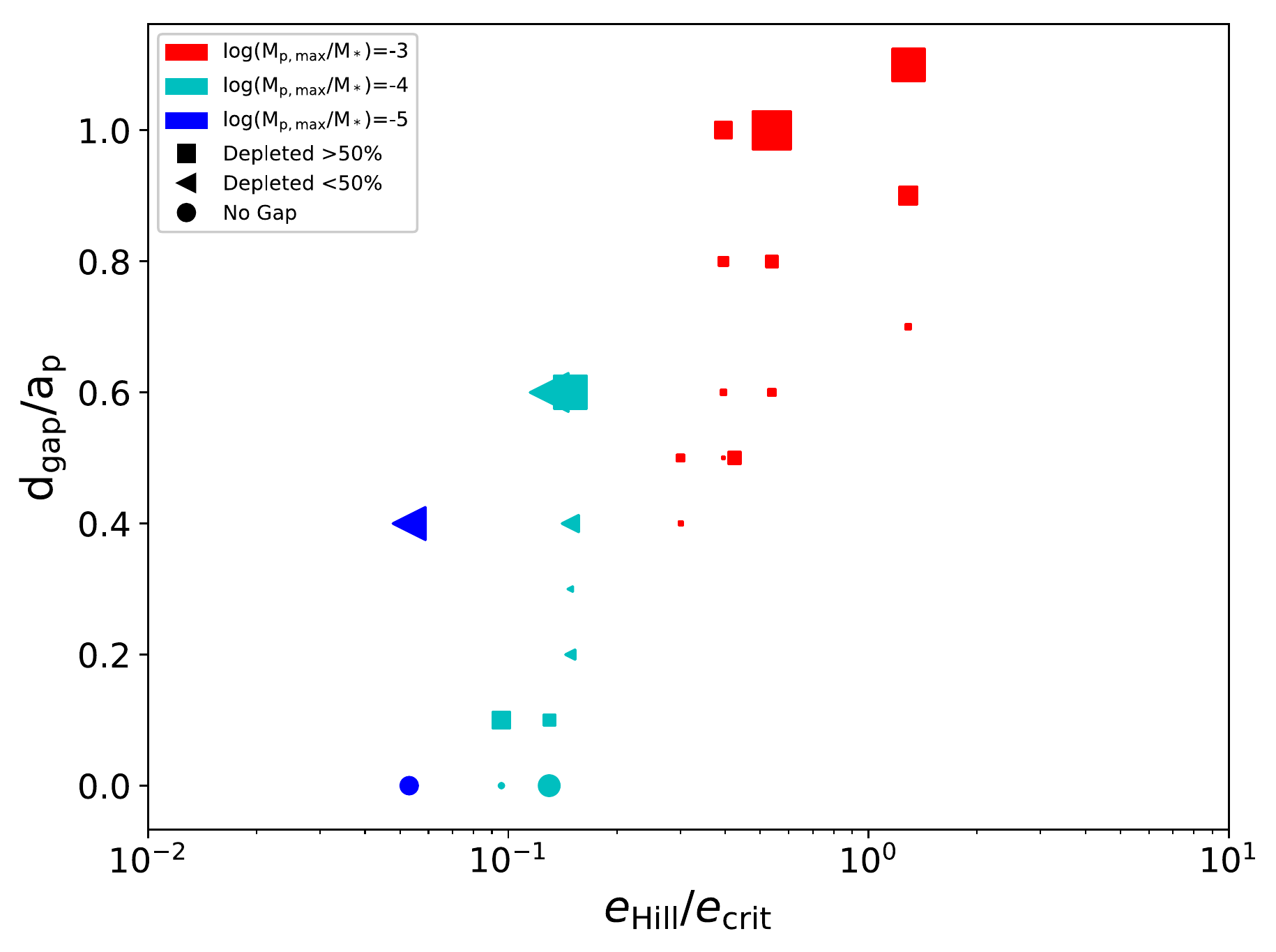}
\caption{Gap width after $10^4$ outer planet orbits versus the ratio of the Hill eccentricity to the initial critical eccentricity required to cross the other planet's encounter zone. Symbol size denotes local disk to total planet mass ratio. Color delineates the highest initial planet mass within the pair. Gap widths are normalized to the initial orbit distance of the outer planet.}
\label{fig:gapw-e}
\end{figure}

The width of gaps formed as planets migrate depends on the planets' migration rates, which, in turn, depend on the degree of sharing of disk material and the disk mass. While the width of gaps formed by non-migrating planets depends on planet mass via the size of a planet's chaotic zone ($\propto(M_p/M_*)^{2/7}$; \citealt{Wisdom:1980}), we find that for migrating planets, gap width is better correlated with $e_{\rm{Hill}}/e_{\rm{crit}}$. The more the highest mass planet within a pair can perturb disk material to cross the encounter zone of the other planet, the wider the gap. There is very little correlation between gap width and $e_{\rm{Hill}}/e_{\rm{crit}}$ with respect to the lowest mass planet within the pair, as expected. The timescale over which gaps form and grow depends on the time needed for the planets to clear disk material as they migrate. As shown in \citet{Morrison:2015}, lower mass, non-migrating planets take longer to clear material. A single planet with $M_p/M_*=10^{-3}$ takes a few hundred orbits, $M_p/M_*=10^{-4}$ takes a few thousand, and $M_p/M_*=10^{-5}$ takes $\sim10^4$ orbits to clear disk material co-planar with a planet. This clearing timescale also increases with disk height, so this co-planar clearing timescale serves as a minimum time to clear disk material for an isolated planet. Figure~\ref{fig:gapw-e} shows the width of gaps from our simulations after $10^4$ orbits as a function of the disk mass and capability of the highest mass planet in the pair to share disk material with the other planet ($e_{\rm{Hill}}/e_{\rm{crit}}$). On these timescales, half a dex higher local disk-to-planet mass ratio of the highest mass planet produces a gap that is about 0.2$a_{\rm{outer}}$ wider for a given initial $e_{\rm{Hill}}/e_{\rm{crit}}$ ratio. Since disk mass primarily determines the migration of the planets, this indicates that the widths of gaps in planetesimal disks depend on the migration of the planets in addition to the initial planetary system architecture, even on these short timescales.

\subsection{Effects of a Planet's Size}
The planet's physical radius relative to its Hill radius affects whether it clears material predominately via accretion or scattering as well as the timescale over which that process occurs \citep{Morrison:2015}. The ratio between a planet's physical radius and its Hill radius weakly depends on planet density and is primarily a function of its orbit distance (see Equation 4 in \citealt{Morrison:2015}). Consequently, a planet's size and starting location should influence its migration particularly for planets that clear nearby disk material predominately by accretion. For the planet mass ranges we consider here, $M_p/M_*=10^{-3}$ planets clear material via scattering, while accretion dominates for $M_p/M_*=10^{-5}$ over typical orbit distances. To investigate the impact of planet size, we performed additional numerical integrations for a subset of our system configurations with at least one planet $M_p/M_*<10^{-3}$. We adopted planet sizes of $R_p=10^{-3}\Rh$ instead of $10^{-2}\Rh$. Note that both planet sizes are comparable to Solar System planets and exoplanets alike (see Figure 1 in \citet{Morrison:2015}). 

Planets filling more of their Hill radius have the potential to accrete more mass, which changes their migration (and gap formation) rates. However, planets with sufficiently high mass to clear nearby disk material via scattering will not have different migration rates for different $R_p$. Additionally, the migration rate of individual planets depends  weakly on planet mass. Initially $M_p/M_*=10^{-5}$ planets typically accreted less than thirty of percent of their mass over $10^5$ orbits for our main $R_p=10^{-2}\Rh$ simulations. Consequently, migration rates in our simulations are not significantly different for different choices of $R_p$ over timescales greater than a few thousand orbits for the same initial planet and disk mass configurations. We do find slight differences in the inner planet's migration rate. This is likely due to the stronger dependence of the inner planet migration rate on the outer planet mass via the sharing ratio between the planets. If the outer planet's mass increases due to accretion, this then impacts the inner planet's migration rate for initially closely spaced planet pairs as investigated here.

Although migration rates change, we do not find different gap widths due to a different $R_P$ except for low mass planet pairs with $M_p/M_*\lesssim10^{-4}$ that contain one $M_p/M_*=10^{-5}$ planet. Therefore, the gap widths arising from low mass migrating planet pairs in our main simulations ($R_p=0.01\Rh$) should be taken as upper limits when applied to gaps in debris disks at astrocentric distances $\gtrsim$5 AU.

\section{Discussion}
\subsection{Inferring Planets from Gaps}
\begin{figure*}
\centering
\includegraphics[scale=0.45]{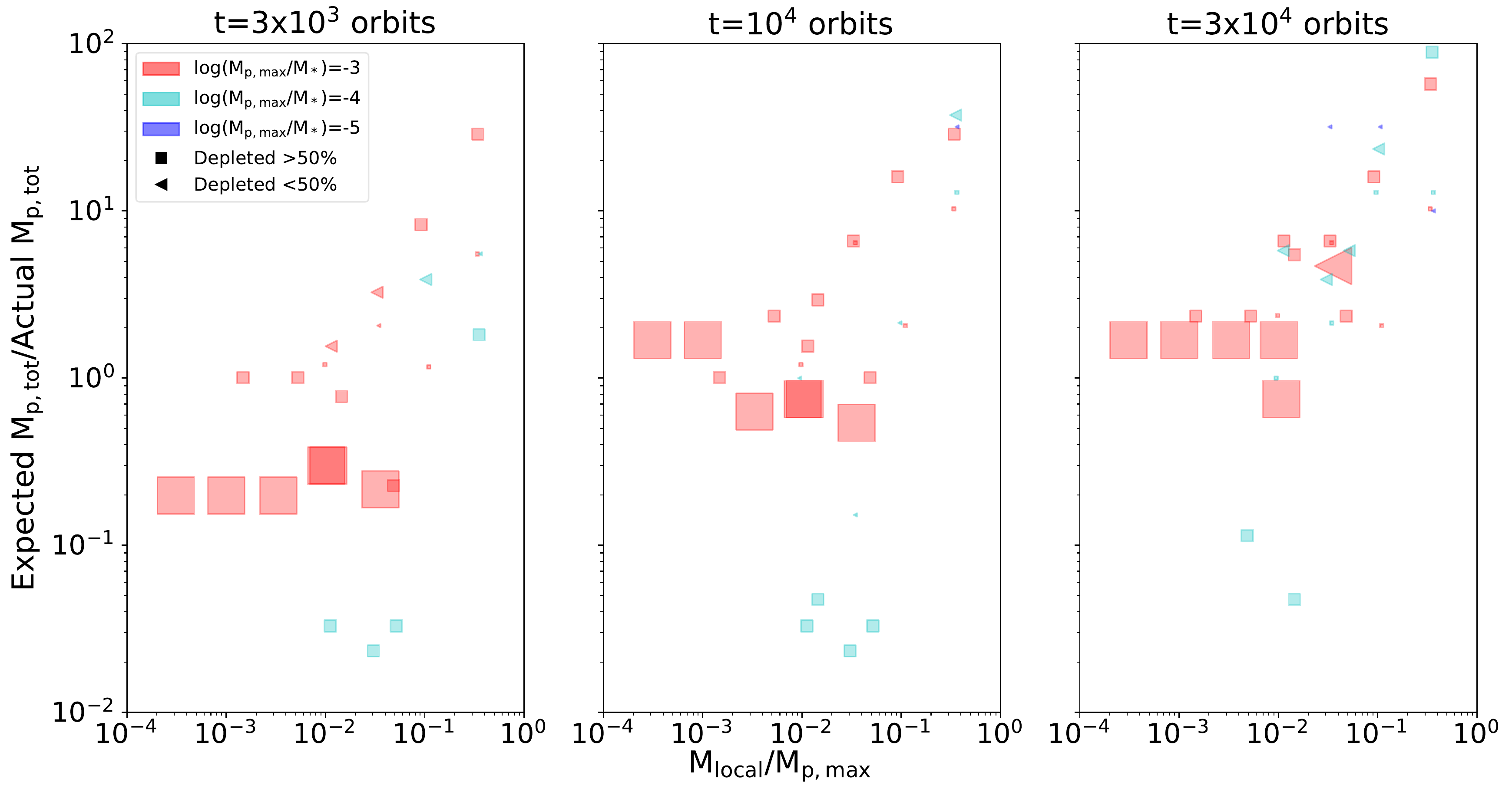}
\caption{A comparison over time of the mass of a planet pair inferred without migration (`expected') to the mass of the migrating pair that produced the given gap width (`actual') as a function of local disk mass-total planet mass ratio. Colors indicate maximum planet mass in each pair and symbol sizes show the relative planet-planet mass ratio from the simulations of migrating planet pairs.}
\label{fig:Mexp}
\end{figure*}

Since our simulations show that closely separated pairs of planets can form gaps in disks as the planets migrate apart, we assess how migration might influence planet masses inferred from gaps. We use analytic estimates for gaps formed by stationary planets to derive the masses of planets that would be inferred from the gaps in our simulations. To calculate the planet masses sufficient to form similar gaps without migration, we use stability constraints of the planets with respect to each other and the boundaries of the gap. First we use a simple dynamical spacing approach: we require that the hypothetical non-migrating planets must be spaced at least 2$\sqrt3$ mutual Hill radii apart, as required to be dynamically stable from orbit crossings with respect to each other \citep{Gladman:1993}. To clear a gap of the given width, we also require that the inner and outer planets must be at least one chaotic zone width away from the inner and outer gap edge, respectively. We compare the maximum planet mass that meet these criteria for equal mass planet pairs to the maximum planet mass in the pair of migrating planets that actually formed the gap in our simulations. In Figure~\ref{fig:Mexp}, we plot the ratio of expected total planet mass in a non-migrating pair to a migrating pair for a given gap versus $M_p/M_*$ for the simulated planets. In \S\ref{s:gapw}, we reported that gap widths correlated with the degree of sharing as measured by $e_{\rm{Hill}}/e_{\rm{crit}}$ with respect to the most massive planet within a migrating pair. As shown in Figure~\ref{fig:Mexp}, the discrepancy between non-migrating and migrating planet masses needed to form a given gap also depends on the highest mass planet within the pair, particularly its local disk-planet mass ratio. Since the gap formed by migrating planets grows with time, we perform this comparison on different timescales. The discrepancy in inferred planet masses between accounting for and neglecting migration can grow up to a couple orders of magnitude within 3$\times10^4$ orbits. This discrepancy was greatest for gaps formed in disks where M$_\mathrm{local}$/M$_\mathrm{p,max}\gtrsim$ 1/10 and with migrating planets that differ from their neighboring planet by 10x in mass.

Because planet migration rates (and changes in spacing between planets) are higher for higher local disk-planet mass ratios, the migration-induced discrepancy at a given time is also higher for higher disk-planet mass ratios. The magnitude of this discrepancy is also both time and planetary system architecture dependent. We can partially account for these dependencies by considering the time it takes non-migrating planets to clear material in their vicinity. Single $M_p/M_*\lesssim10^{-4}$ planets likely could not have cleared gaps on their own within 3,000 orbits unless migrating appreciably in high mass disks and/or in the presence of nearby additional planets (as shown in the leftmost plot, Figure~\ref{fig:Mexp}). 

It seems counterintuitive that the migrating pairs produce underestimates, but this is because the simple analytic approach used previously does not account for the time it takes multiple planets to clear a gap. Using Equation 6 from \citet{Shannon:2016}, we estimate the minimum mass of non-migrating planets within multi-planet systems required to clear gaps within a given time. If the lowest mass planet within the pair falls under this minimum mass limit from \citet{Shannon:2016} for a given timescale, then the analytical planet mass expectations from assuming non-migrating planets will underestimate both the total mass in planets and the highest planet mass within a migrating pair (symbol color in Figure~\ref{fig:Mexp}). For the range of planet and disk masses we consider in this study, this occurs for planet pairs containing a $M_p/M_*=10^{-5}$ planet embedded in disks with local disk-planet mass ratios of less than 0.1; relevant for young, debris disk systems with potential planet signatures at large orbit distances. 

In low mass disks where $M_{\rm{local}}/M_{\rm{p}}<0.01$, the planets do not migrate appreciably for the planet masses simulated here, so the no-migration inferred planet mass prediction more closely agrees with the migrating planet mass beyond $\sim$10x the co-planar clearing timescale of the highest mass planet in the pair. For the timescales included in Figure~\ref{fig:Mexp}, that corresponds to pairs containing $M_p/M_*=10^{-3}$ planets. Inferred non-migrating planet masses from gaps formed by migrating planet pairs with very unequal planet masses tend to correctly recover the maximum planet mass, but over estimate the total mass. For example, as shown in Figure~\ref{fig:Mexp}, a system with a migrating $M_p/M_*=10^{-3}$ and $M_p/M_*=10^{-5}$ planet produces a gap over a wide range of disk masses nearly as wide as in a system with two non-migrating $M_p/M_*=10^{-3}$ planets. 

One final difference not accounted for in Figure~\ref{fig:Mexp} is the typical depletion in gaps. While divergently migrating planets typically create large gaps at fixed planet mass, disk material may remain in the in between them if the sharing ratio is less than 1. For the initial planet separation we examine here (initial period ratio of 1.84), this occurred in planet pairs containing a $M_p/M_*=10^{-5}$ planet (see Figure~\ref{fig:gapdep} for some examples). Additionally, as planets migrate through the disk and continue to scatter disk particles, they may transfer disk material from one side of the planet's orbit into the gap. In contrast, two non-migrating planets would not continually scatter disk material into the gap to the same degree over time. 

To keep the simulation parameter space manageable, we have restricted our current study to a single initial semi-major axis ratio between the planets. Our initial conditions were chosen to probe the regime in which planet-planet interaction would be strongest, while still allowing planets to maintain stable orbits at all mass ratios. More widely separated pairs with negligible sharing ratios will initially mimic the single planet results.  Because migration rates increase with semi-major axis for typical disk surface density profiles, widely separated pairs should also tend to undergo convergent migration, driving them back into the regime we have explored here.

\subsection{Disk Masses and Planet Starting Locations}\label{S:ictranslate}

Since the distance that planets migrate depends on $M_{\rm{local}}/M_{\rm{p}}$, we consider the implications of this study in the context of observed debris disk systems and disk profiles relevant for planet formation. Masses of debris disks are not well constrained; observations in the infrared and even at sub-mm probe debris particle sizes smaller than the planetesimal masses driving planet migration. By extrapolating size distributions up to the largest sizes that participate in collisional cascades, observed large cold debris disks (like $\beta$ Pic or HR 8799) are thought to contain $\sim100M_\oplus$ in mass, but this estimate relies on uncertain properties of collisionally evolved debris disks, such as the height of the disk and maximum planetesimal size \citep{Moore:2013}. The degree to which current gapped debris disks reflect their progenitor disk of solids following protoplanetary disk dispersal is also unknown. From a modeling perspective, in the `Nice' models, the planetesimal-driven migration of Uranus and Neptune laid the foundation for outer planet orbit instabilities determining aspects of the modern Solar System architecture. The migration of Uranus and Neptune in those models required a 30-50 $M_\oplus$ disk 10s of AU in extent \citep{Tsiganis:2005}, or $M_{\rm{local}}/M_{\rm{p}}\sim0.1$ as parameterized in this study.

\begin{figure}
\centering
\includegraphics[width=\columnwidth]{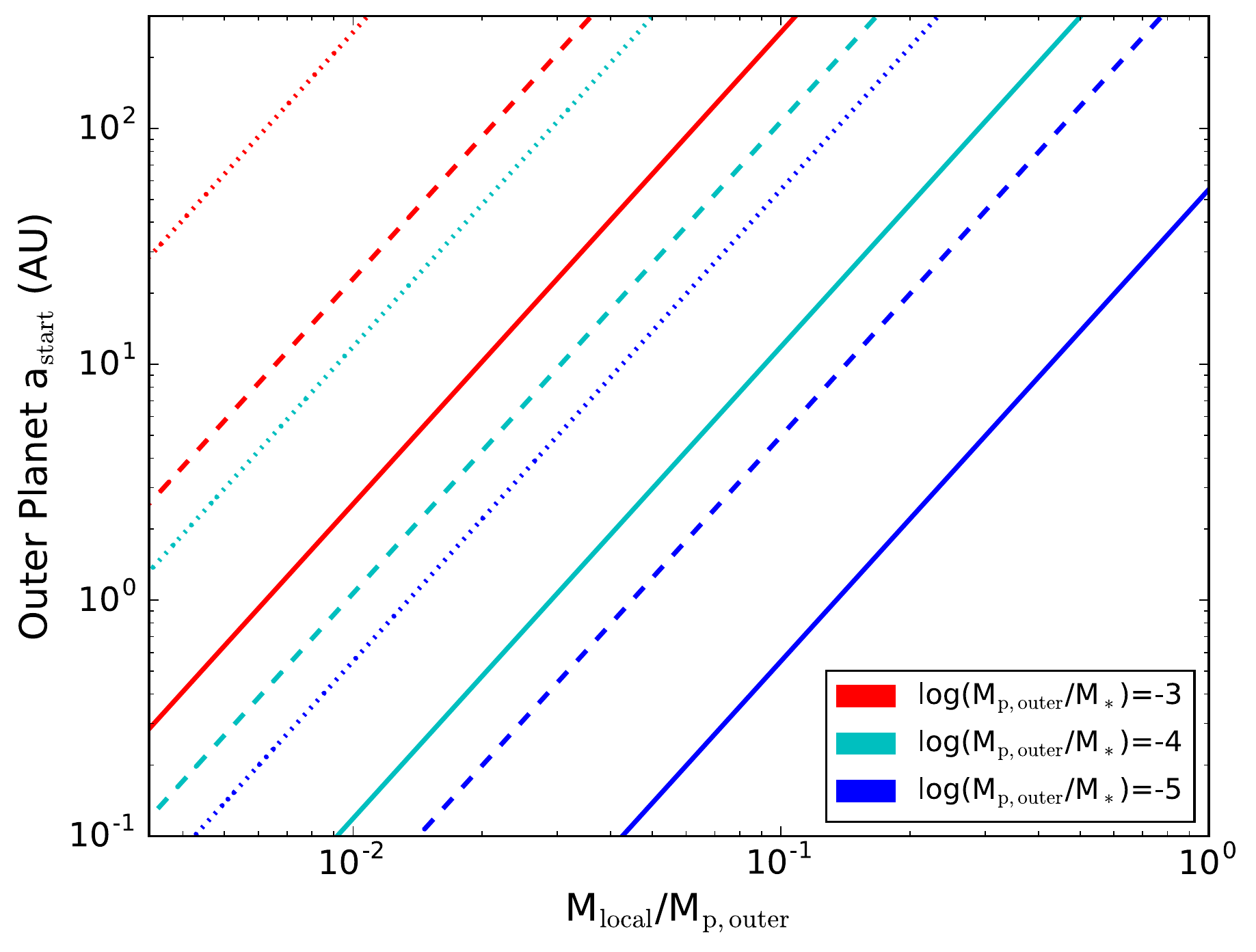}
\caption{Initial semimajor axis of the outer planet for a given planet mass and $M_{\rm{local}}/M_p$ within a disk of solids following a disk profile that is 0.3, 1, or 3 times the solid portion of a minimum mass solar nebula profile (dotted, dashed, and solid line, respectively), adopting the solid disk is 0.01 times the mass of the original protoplanetary disk.}
\label{fig:MMSNastart}
\end{figure}

We compare the local disk conditions for which we observe divergent planet migration and gap formation in our simulations to potential progenitor solid disks for debris disks. We scale these by the minimum mass solar nebula surface density profile at a given starting distance. The profile we consider is 
\begin{equation}
\Sigma_{\rm{MMSN}}=1700 \left(\frac{a}{\text{AU}}\right)^{-1.5} \text{g/cm$^2$}
\label{eq:MMSN}
\end{equation}
from \citet{Hayashi:1981}. As the initial disk conditions for divergent planetesimal-driven migration, we assume a gas-to-dust ratio of 100:1 and adopt that the solid portion of this disk, $\Sigma_{\rm{MMSN, solid}}$, is 0.01$\Sigma_{\rm{MMSN}}$. In disks with mass profiles 0.3 to 3$\Sigma_{\rm{MMSN, solid}}$, the equivalent starting locations of the outer planet for a given $M_{\rm{local}}/M_{\rm{p}}$ are shown in Figure~\ref{fig:MMSNastart}.

\begin{figure}
\centering
\includegraphics[width=\columnwidth]{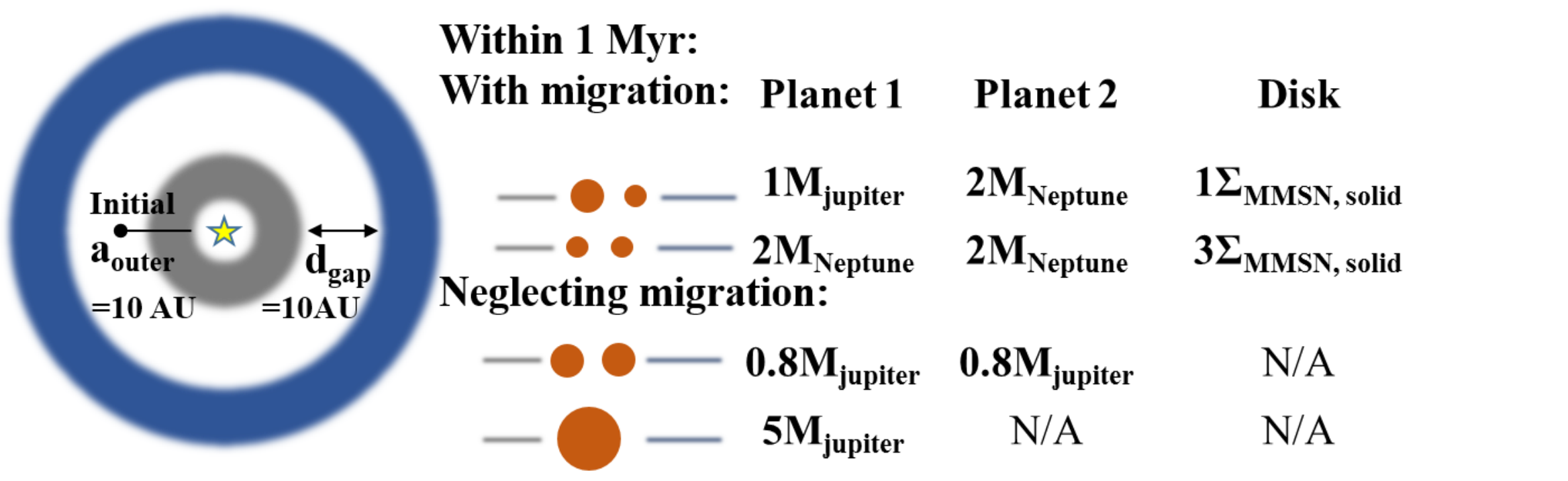}
\caption{Example planet configurations inferred from a 10 AU wide gap for a 1$M_\odot$ host star, including responsible migrating planet pairs from our simulations and analytic estimates from neglecting planetesimal-driven orbit migration of planets. The initial location of the outer migrating planet is at 10AU.}
\label{fig:plconfigcartoon}
\end{figure}

Using these outer planet starting distances, we then translate the migration and gap formation timescale from our models into years. For outer planets that start migrating at distances of $\sim$10 AU, planet pairs in our simulations produced $\sim$ 10 AU wide gaps in less than 3 Myr if they contained a $M_p/M_*=10^{-3}$ planet or two $M_p/M_*=10^{-4}$ planets in disks down to 0.3$\Sigma_{\rm{MMSN, solid}}$. Examples of inferred planet configurations at 1 Myr for a hypothetical disk and gap configuration are shown in Figure~\ref{fig:plconfigcartoon}. Migrating planets could plausibly produce observed debris disk gaps yet be low enough mass to elude current detection. Moreover, the inferred planet masses when neglecting migration for such gaps could be expected to be observable by direct imaging surveys for young, nearby systems.

 \section{Conclusions}
Motivated by the ongoing observational characterization of gapped debris disk systems, in this study we examined what planet mass and disk mass combinations allow pairs of planets to divergently migrate and produce gaps in planetesimal disks. From our simulations, we find that divergently migrating pairs of planets in planetesimal disks can form gaps as wide as a couple times the outer planet's orbital separation within a few Myr at 10s of AU. Pairs of divergently migrating planets can form gaps in planetesimal disks $\sim$0.1\% of the minimum mass solar nebula with a lower total mass in planets than would be expected for a gap carved with non-migrating planets. Moreover, these migrating, gap opening planets are less than the typical $\sim$few $M_{\rm{Jupiter}}$ lower detection limits of current direct imaging surveys \citep{Bowler:2016}. As demonstrated here, perhaps forming gaps with lower mass, migrating planets helps reconcile the lack of direct imaging detections for gapped debris disks in which estimates neglecting migration would predict the presence of high mass planets. 

In this study we also show that the migration of planets in debris disks can be strongly affected by the presence of nearby planets. The migration rates of individual planets within planet pairs are slower than for a single isolated planet especially if planets exchange scattered populations of planetesimals at early times and deplete the disk between planets. This typically results in the divergent migration of initially closely separated planets. The degree of sharing between the pair of planets introduces variation in an individual planet's migration rate for a given planet and disk mass. As a result, gap widths depend on both the disk mass and indirectly on the planet mass through its sharing ratio with its neighboring planet. At fixed initial planet-planet separation, the widest gaps form when the highest mass planet in the system has a large sharing ratio. The local disk to planet mass ratio also influences the width of gaps to within a factor of a few for a given sharing ratio. 

The migration of the planets as well as the gaps they produce are not simply determined by the disk that they are migrating through. It also depends on the planetary system architecture itself, so inferring the properties of planets from gaps in debris disks can be challenging. We have, however, placed some constraints on when planet migration should be considered. Pairs of super-Earths, except in high mass disks, do not produce sizable gaps fast enough to be responsible for gaps in young ($\sim$20 Myr) debris disks. Wide gaps (R$_{\rm outer}$/R$_{\rm inner}\sim$10) in young debris disks, as exhibited in some observed systems, still would require more than two planets to produce clearing over these large distance scales over the system's lifetime. To continue to develop realistic expectations of possible planets in disk systems, divergent planet migration in the evolution of young debris disks warrants further study.

\section*{Acknowledgments}
We thank the anonymous reviewer whose comments and suggestions greatly improved this manuscript. This research made use of the NASA Astrophysics Data System Bibliographic Services and the computational facilities of the University of Arizona Theoretical Astrophysics Program. SJM is supported in part by The Center for Exoplanets and Habitable Worlds at the Pennsylvania State University. The Center for Exoplanets and Habitable Worlds is supported by the Pennsylvania State University, the Eberly College of Science, and the Pennsylvania Space Grant Consortium. KMK is supported in part by the National Science Foundation under Grant No. AST-1410174.

\bibliographystyle{mnras}

\bsp	
\label{lastpage}
\end{document}